\documentclass{llncs}

\makeatletter
\def\@Thm#1#2#3{\topsep 7\p@ \@plus2\p@ \@minus4\p@
\@ifnextchar[{\@Ythm{#1}{#2}{#3}}{\@Xthm{#1}{#2}{#3}}}
\def\@spthm#1#2#3#4{\topsep 3\p@ \@plus1\p@ \@minus2\p@
\refstepcounter{#1}%
\@ifnextchar[{\@spythm{#1}{#2}{#3}{#4}}{\@spxthm{#1}{#2}{#3}{#4}}}
\renewcommand\section{\@startsection{section}{1}{\z@}%
                       {-10\p@ \@plus -2\p@ \@minus -2\p@}%
                       {4\p@ \@plus 2\p@ \@minus 2\p@}%
                       {\normalfont\large\bfseries\boldmath
                        \rightskip=\z@ \@plus 8em\pretolerance=10000 }}
\renewcommand\paragraph{\@startsection{paragraph}{4}{\z@}%
                       {-3\p@ \@plus -2\p@ \@minus -2\p@}%
                       {-0.5em \@plus -0.22em \@minus -0.1em}%
                       {\normalfont\normalsize\itshape}}
\makeatother

\usepackage[utf8]{inputenc}
\usepackage[T1]{fontenc}
\usepackage{microtype}
\DeclareMathAlphabet{\mathcal}{OMS}{cmsy}{m}{n}
\usepackage{caption}
\usepackage{tikz}

\usepackage{color}

\usepackage{amssymb, amsmath,  graphicx}
\usepackage{url}
\usepackage{xspace}
\usepackage{paralist}
\usepackage{graphicx}
\usepackage{tabularx}
\usepackage{booktabs}
\usepackage{needspace}

\newcommand{\pr}{\mathrm{Pr}}

\DeclareMathOperator*{\argmax}{arg\,max}

\newcommand{\quality}[1]{#1}

\usepackage[pdfborder={0 0 0}, plainpages, pdfpagelabels=false, pdfstartview=FitH,hidelinks]{hyperref}

\begin{document}

\title{Uncertainty in Crowd Data Sourcing\\
under Structural Constraints 
}

\author{Antoine Amarilli\inst{1} \and Yael Amsterdamer\inst{2} \and Tova
Milo\inst{2}}

\institute{Institut Mines--Télécom; Télécom ParisTech; CNRS LTCI, Paris, France
  \and
Tel Aviv University, Tel Aviv, Israel
}

\maketitle
\begin{abstract}
  \vspace{-1.2em}
  Applications extracting data from crowdsourcing platforms must deal with the
  \emph{uncertainty} of crowd answers in two different ways: first, by deriving
  estimates of the correct value from the answers; second, by choosing crowd
  questions whose answers are expected to minimize this uncertainty relative to
  the overall data collection goal. Such problems are already challenging when we assume that
  questions are unrelated and answers are independent, but they are even
  more complicated when we assume that the unknown values follow hard
  \emph{structural constraints} (such as monotonicity).

In this vision paper, we examine how to formally address this issue with an approach inspired by~\cite{amsterdamer2013crowd}. We describe a generalized setting where we model constraints as linear inequalities, and use them to guide the choice of crowd questions and the processing of answers.
We present the main challenges arising in this setting, and propose directions to solve
them.

\end{abstract}

\section{Introduction}
\label{sec:intro}
\emph{Crowd data sourcing} leverages human knowledge to obtain information which does not exist
in conventional databases. This may be done by posing targeted questions to crowd users, through conventional crowdsourcing platforms such as Amazon Mechanical Turk~\cite{mturk}. Contrary to many works that use the crowd as a \emph{means} to perform different tasks, here the crowd serves as a \emph{source} of information.

Many challenges arise when using the crowd as a data source. First, human
answers have a high latency and are usually provided against some (monetary) compensation, so we must minimize the number of posed questions.
Second, answers collected from the crowd may be erroneous and noisy, so we must control and improve answer quality, e.g., pose the same question to multiple workers.

A vast body of research has tackled these issues for
various data procurement tasks (e.g.,~\cite{amarilli2014complexity,amsterdamer2013crowd,parameswaran2012crowdscreen,parameswaran2011human,trushkowsky2013crowdsourced,yang2013incentive}). For example, \cite{parameswaran2012crowdscreen} studied the number of answers that must be obtained to reach sufficient confidence in the final answer of a given Boolean question, and mentions the problem of deciding, when there are several questions to answer, which is the \emph{next best question} to ask the crowd.
In different situations~\cite{amsterdamer2013crowd,yang2013incentive}, this selection of questions is
performed by comparing the
expected contribution of the answers to some data acquisition goal. However, in such situations, the answers to the various questions are independent, so that we can choose the next best question by looking at each question in isolation.

In this paper, we study the problem of collecting numerical values from the
crowd under \emph{hard a-priori constraints} on the final answers, caused by inherent data dependencies.

For instance, suppose that we have devised a lossy compression
algorithm for e.g.\ music files, and that we wish to estimate the average quality rating of different compression ratios in a user population.
We can ask a few random crowd workers to evaluate the quality $\quality{q_1}, \ldots, \quality{q_n}$ for each of $n$ compression ratios of \emph{increasing} lossiness.
The quality ratings of any given person are not independent: we can assume that every person will consider $\quality{q_1}$ (the quality of the least lossy compression) to be at least as high as $\quality{q_2}$, and so on. Consequently, the average $\quality{q_1}$ in the entire population is higher than $\quality{q_2}$, and so on. However, the quality $\quality{q_1}$ for \emph{some} people might be lower than $\quality{q_2}$ for others; hence, by asking random workers we may obtain an estimation of the quality ratings that is not perfectly monotone. This use case will be our running example throughout the paper.

As another example, consider the estimation of the price that people are willing to pay for varying combined deals. In fields such as auction study in game theory~\cite{parkes2000iterative}, it is customary to assume that the price function for each user is monotone, i.e., adding products cannot decrease the deal price. For instance, we know that in the entire population, the average value of a flight and hotel cannot be lower than that of the flight alone. But again, if we sample different users for each deal, we may obtain a non-monotone estimation for the average price.

A similar problem occurs in~\cite{amsterdamer2013crowd}, where the crowd is used to estimate the frequency of patterns in user habits. While these frequencies are dependent for patterns with overlapping activities (e.g., if someone never swims, they also never swim and dive), such dependencies are not accounted for in~\cite{amsterdamer2013crowd}. In general, existing work on crowd data sourcing has mostly ignored the problem
of uncertainty when dealing with dependent
questions~\cite{amarilli2014complexity,parameswaran2011human}. There are works that deal in a non-trivial way with the
interaction between uncertainty and
dependency~\cite{karp2007noisy,triantaphyllou2010data}, but they assume that the
individual outcomes observed are Boolean and not numeric like in the present
paper.

We consider here two important problems that arise in the context of dependent crowd questions.
First, can we \emph{improve the variable estimation} by taking dependencies into account? For instance, in our running example, if we estimate that the quality $\quality{q_1}$ is lower than $\quality{q_2}$ (which contradicts our monotonicity assumption), we may attempt to correct our estimation by increasing $\quality{q_1}$ and/or decreasing $\quality{q_2}$; or, to begin with, we can only consider estimations that comply with our monotonicity requirement. What is the right way to enforce this monotonicity, and how does it increase the quality of our estimation?

Second, we use the dependencies to \emph{reduce the number of questions} posed to the crowd. For instance, if we estimate that the average quality rating $\quality{q_1}$ and $\quality{q_4}$ of the compression ratios 1 and~4 are both~6 out of~10, we do not need to ask people about $\quality{q_2}$ and $\quality{q_3}$. Or, as another example, if we wish to find the lossiest compression with rating at least~6 out of 10, and we estimate that $\quality{q_5}$ and $\quality{q_{10}}$ are~8 and~3 respectively, we can interpolate $\quality{q_6}, \ldots, \quality{q_9}$ (using monotonicity) and estimate that the rating with value closest to~6 is most likely $\quality{q_7}$.

\paragraph{Paper structure.}
We first give a formal definition of the considered problem in
Section~\ref{sec:problem}. We next present in Section~\ref{sec:without} a
general scheme to solve the problem in the absence of dependencies, and
turn in Section~\ref{sec:with} to how dependencies should be handled. For
numerous dependent variables,
interpolating samples is crucial: we discuss this in Section~\ref{sec:interpolation}.
Last, we conclude in Section~\ref{sec:conclusion}.

\section{Problem statement}
\label{sec:problem}
We wish to learn $n$ numerical values $\boldsymbol{\mu} = (\mu_1, \ldots,
\mu_n)$ from the crowd. We model the \emph{distribution of crowd answers} to questions about these values using $n$ \emph{random variables}
$X_1, \ldots, X_n$. We assume that the \emph{mean} of $X_i$ is $\mu_i$ for every $i$.\footnote{This assumption holds when we are interested in the \emph{average} crowd answer, e.g., the average rating for a compression quality; and in the many cases where the errors of worker answers tend to cancel out so that the average is close to the truth~\cite{amsterdamer2013crowd}.}

We further assume that
$\boldsymbol{\mu}$ satisfies a certain known set of \emph{linear inequalities},
represented as a matrix $E$ of reals such that $E \cdot \boldsymbol{\mu} \leq
(0)$, where $(0)$ is the zero vector and $\cdot$ denotes the product of matrix
$E$ and vector $\boldsymbol{\mu}$. We assume that the inequalities $E$ are
feasible, namely, that there is some vector $\mathbf{e}$ satisfying $E$.

\begin{example}
\label{exa:ineq}
In our running example, the random variables $Q_1, \ldots, Q_n$, with unknown
means $q_1, \ldots, q_n$, denote the ratings obtained for the compression
ratios. The inequalities represent a decreasing order: $q_2 - q_1 \leq 0$,
$q_3 - q_2 \leq 0$, etc.
\end{example}

We consider a known \emph{loss function} $L_{\boldsymbol{\mu}}$ which associates to
a prediction $\mathbf{v}$ for the unknown values $\boldsymbol{\mu}$ some nonnegative
value $L_{\boldsymbol{\mu}}(\mathbf{v})$. We assume that $L_{\boldsymbol{\mu}}$ can be
written as the sum of nonnegative functions $L^i_{\mu_i}$, that is, the error
function for all values is the sum of the errors of individual values. We
require that for all $i$, $L^i_{\mu_i}(\mu_i) = 0$ and $L^i_{\mu_i}(x) \leq
L^i_{\mu_i}(y)$ for all $\mu_i \leq x \leq y$ and $y \leq x \leq \mu_i$. (In
other words, the loss is $0$ for the correct value, and increases with the
absolute error.)

\begin{example}
\label{exa:loss}
The loss function depends on the target application. For compression
ratios, if our task is to find which is the lossiest compression with rating at
least $6$, a reasonable loss function for all variables is the \emph{threshold}
loss $L_{\mu, \tau}$ with $\tau = 6$. The value $L_{\mu, \tau}(x)$ is defined to
be $1$ if the $x$ and $\mu$ are miscategorized with respect to threshold $\tau$
(formally, $\mu < \tau < x$ or $x < \tau < \mu$) and $0$
otherwise. The overall loss function is the sum of the $L_{q_i,
\tau}$ which counts the number of ratios that are miscategorized with respect to
the threshold $\tau = 6$.
\end{example}

For any $i$, we can obtain a sample of variable $X_i$ (we say that we
\emph{sample} $X_i$ or \emph{draw} $X_i$) by asking the corresponding question
to a random crowd worker; we assume that all draws are independent both between
variables and between two draws of the same variable.
Our goal is to choose draws carefully and, based on the obtained samples, try to provide a prediction $\mathbf{v}$
which minimizes $L_{\boldsymbol{\mu}}(\mathbf{v})$:
we phrase this in a
\emph{fixed-budget} formulation, namely minimize
$L_{\boldsymbol{\mu}}(\mathbf{v})$ in expectation after a fixed number of
samples.

\begin{example}
In the running example, sampling the variable $Q_i$ is achieved by providing a
random crowd user with a sound sample compressed with ratio $i$ and
asking for a rating for this sample. The overall objective is to choose
the right ratios for which to request more ratings, in order to minimize the
number of average quality ratings that are miscategorized with respect to $\tau
= 6$.
\end{example}

We next review the problem of minimizing the loss by choosing the ``right'' questions.
We first study an approach for a simplified setting where there are no order constraints on the estimated values, before we consider the general case.

\section{Without order constraints}
\label{sec:without}
Let us present a general scheme inspired by~\cite{amsterdamer2013crowd} for the
case with no order constraints, before we extend it to order constraints in the
next section.

With no constraints, as the variables are independent and the loss
is the sum of the individual losses of variables, our goal is to
find which one of the variables is such that one more sample for it would yield the largest loss reduction.
Hence, we first focus on an individual variable $X_i$ to
describe how we predict its mean value $v_i$ from the samples $S_i$ observed for this variable, and how we
estimate the loss reduction that we may achieve by taking one more sample.

\paragraph{Estimating the parameter.}
Our approach for a variable $X$ given a set $S$ of
samples of this variable is to fit a model for $X$ from the family of normal distributions, as they are a simple
and general way to represent real-life data. Denote by $\Theta = \mathbb{R} \times \mathbb{R}_+$ the \emph{parameter space}, such that every
$\theta \in \Theta$, with $\theta = (\mu, \sigma^2)$, represents the normal distribution $\mathcal{N}(\mu, \sigma^2)$ with mean $\mu$ and variance $\sigma^2$. Denote by $\pr_{\theta}$ the probability density function of this distribution.

As the samples $S$ of $X$ are assumed to be independent, we can define the
probability of $S$ according to $\mathcal{N}(\theta)$ as the product of
$\pr_{\theta}(s_i)$ for all $s_i \in S$.
The \emph{likelihood function} $\mathcal{L}_S : \Theta
\rightarrow [0, 1]$ is then simply defined as $\mathcal{L}_S(\theta) =
\pr_{\theta}(S)$:
it describes, \emph{as a function of $\theta$}, the
probability\footnote{Note that likelihood cannot, however, be seen as
a probability distribution on $\Theta$.} of the sample under $\theta$.

Our way to fit a normal distribution to the random variable $X$ is then
the standard method of choosing the \emph{maximum likelihood estimator} (MLE):\vspace{-2mm}
\[\vspace{-2mm}
  \widehat{\theta} = \argmax_{\theta \in \Theta} \mathcal{L}_S(\theta)
\]
In the case of normal distributions, it is easily checked that we have $\widehat{\theta}
= (\widehat{\mu}, \widehat{\sigma}^2)$, where $\widehat{\mu}$ and
$\widehat{\sigma}^2$ are the \emph{sample mean} and \emph{sample variance}
defined by:\vspace{-1mm}
\begin{align*}
  \widehat{\mu} = & {1 \over |S|} \sum_i s_i & \widehat{\sigma}^2 = & {1 \over |S|} \sum_i (s_i - \widehat{\mu})^2
\end{align*}
\vspace{-2mm}Hence, we take $v = \widehat{\mu}$ as our current guess of the mean of variable $X$.

\begin{example}
  Assume that we ask $3$ users to evaluate sound samples compressed with
  ratio $3$, and obtain the grades $3$, $5$, and $7$. This means
  that our sample mean and variance for variable $Q_3$ are respectively
  $\widehat{\mu_3} = 5$ and
  $\widehat{\sigma_3}^2 = 8/3$.
\end{example}

\paragraph{Estimating the error.}

How to estimate the loss of our
prediction $\widehat{\mu}$?
Because the true value is unknown, we estimate the loss by assuming that our
current guess $\widehat{\theta}$ is correct, and finding out what its expected error is. We do this by examining the range of samples that could have been obtained instead of $S$ under the assumed distribution and computing the loss of the MLE obtained from them.

By the central limit theorem, the distribution of the mean of $N$
samples of $\mathcal{N}(\widehat{\mu}, \widehat{\sigma}^2)$ can be approximated
by $\mathcal{N}(\widehat{\mu}, \widehat{\sigma}^2/N)$.
Hence, under the assumption that $\widehat{\theta}$ is correct, we can define
the average error obtained through the MLE method from $|S|$ samples, as follows:
\vspace{-2mm}
\[\vspace{-2mm}
  E(\widehat{\theta}, |S|) = \int_{x \in \mathbb{R}}
  \pr_{\left(\widehat{\mu}, \widehat{\sigma}^2/|S|\right)}(x)
  L_{\widehat{\mu}}(x)\,\mathrm{d}x
\]
This integral can be numerically approximated by sampling.

\begin{example}
  The estimated error for $Q_3$ under the samples $S_3$ of the previous example is
  the probability that the sample mean, distributed according to $\mathcal{N}(5,
  (8/3) \cdot (1/3))$, is above threshold $\tau = 6$ (as the loss is then $1$, and is $0$
  otherwise, relative to our estimate $\widehat{\mu_3} = 5$). Numerically we
  have $E(\widehat{\theta}_3, |S_3|) = 0.144$.
\end{example}

\paragraph{Estimating the error decrease.}
Now that we can estimate the parameter of a distribution from the samples, and
the expected error according to this parameter, we can easily devise an
estimation of how this error may decrease when an additional sample is requested
from variable $X$.

Let us assume that we obtain a new sample of $X$ with value $x$, and call $S'$
the $|S|+1$ samples obtained by adding $x$ to $S$. Call $\widehat{\theta}'$ the
MLE obtained by maximizing $\mathcal{L}_{S'}$, and define the \emph{error
decrease} as $D(S, x) = E(\widehat{\theta}, |S|) - E(\widehat{\theta}', |S|+1)$.
This gives us an estimation of how error decreases for one more sample with
value~$x$. Of course, we cannot know if we would indeed obtain value $x$,
but we can compute its probability according to our current
hypothesis for the underlying distribution, namely $\mathcal{N}(\widehat{\mu},
\widehat{\sigma}^2)$. We therefore define the \emph{expected error decrease}:
\vspace{-2mm}
\[\vspace{-2mm}
  D(S) = \int_x \pr_{\widehat{\theta}}(x) D(S, x) \mathrm{d}x
\]
This is our estimate of the expected loss reduction when sampling this
variable.

\begin{example}
  If we obtain one additional sample of $5$ for $Q_3$ (yielding $S_3'$), the
  estimated mean $\widehat{\mu}_3 = 5$ is unchanged, but the estimated variance
  decreases to $2$ so the estimated error under the new MLE
  $\widehat{\theta}_3'$ becomes $E(\widehat{\theta}_3', |S_3'|) = 0.079$.
  We estimate the expected error decrease by averaging the decrease under
  possible additional samples drawn from our estimated distribution
  $\mathcal{N}(\widehat{\theta}_3)$ for $Q_3$.
\end{example}

\paragraph{Multiple variables.}
With the above method, we can compute the expected
error decrease of each variable, and sample the one whose expected
error decrease is highest. It is easy to see that this greedy approach is optimal in terms
of reducing the expected error over any fixed number of requests, as samples
for one variable do not change the estimated parameter or expected
error of other variables.

\section{With order constraints}
\label{sec:with}
Under order constraints, the problem is more challenging.
Though the loss function remains a sum of loss
functions over individual parameters, it is not possible anymore to manage variables
separately, because information obtained for one variable gives us additional
information about the other variables.
Reconsidering our running example, under the objective of identifying the
  lossiest compression ratio with average quality at least $\tau$, it makes
  little sense to consider the results of every variable independently, and we
  should examine the results globally to locate
  where the decreasing sequence of qualities intersects the threshold $\tau$.
The challenge is how to formalize such a global strategy, under general constraints.

To this end, we propose a greedy strategy inspired by that of the
previous section, but integrating the order constraints and considering the
variables globally rather than in isolation.
Because additional samples on one variable give us information about other
variables, such a greedy approach is no longer guaranteed to be optimal
over multiple draws.
Because of space constraints, we
only sketch the principles of our initial approach; we plan to study this further and
examine possible alternative approaches in future work.

We consider the parameter space $\Theta = (\mathbb{R} \times
\mathbb{R}_+)^n$, covering all parameters of all random variables simultaneously, and
we define the likelihood of $\boldsymbol{\theta} \in \Theta$ (with $\theta_i =
(\mu_i, \sigma_i^2)$) as a function of
$\mathcal{S} = (S_1, \ldots, S_n)$, the set of all samples for all variables,
using the fact that all draws are still independent.
We exclude parameters which violate order
constraints by defining the likelihood as follows:\vspace{-1mm}
\[\vspace{-1mm}
  \mathcal{L}_{\mathcal{S}}(\boldsymbol{\theta}) = \left\lbrace\begin{array}{ll}
    \prod_{i} \prod_{s \in S_i} \pr_{\theta_i}(s) & \mbox{if~}E \cdot \boldsymbol{\theta} \leq (0)\\
    0 & \mbox{otherwise}\end{array}\right.
\]

The main problem is now to determine the maximum likelihood estimator for
$\boldsymbol{\theta}$ by maximizing this expression.
We next propose a possible approach to the problem, and the challenges yet to be resolved.

\paragraph{Estimating the means.}
We propose to maximize the expression as a function of the means
$\boldsymbol{\mu}$, while making the assumption that the variances are the
sample variances $\boldsymbol{\widehat{\sigma}}$ for every individual variable.
Under this approximation, the maximization problem can be rewritten as
maximizing a quadratic expression with a positive definite matrix under the
inequalities $E$. Such a problem is tractable~\cite{kozlov1980polynomial}, so we
can solve it and obtain a set of candidate means $\mathbf{v}$ for the underlying
distributions. Technical details are omitted for lack of space.

\begin{example}
  Assume that we have obtained the same number of samples for $Q_1$, $Q_2$ and
  $Q_3$, that their sample variances are equal ($\widehat{\sigma_1} =
  \widehat{\sigma_2} = \widehat{\sigma_3}$), and that the sample means are
  $\widehat{\mu_1} = 9$, $\widehat{\mu_2} = 7$, and $\widehat{\mu_3} = 8$.
  Observe that we have $\widehat{\mu_2} < \widehat{\mu_3}$ even though we know that
  $q_3 \leq q_2$. In this specific setting, our estimation of the means is the
  solution $\mathbf{v}$ of a
  quadratic programming problem amounting to minimizing the sum of
  squares $\sum_i (v_i - \widehat{\mu_i})^2$ subject to the inequalities: its solution is $v_1 = 9$, $v_2 = 7.5$, and $v_3 = 7.5$.
\end{example}

\paragraph{Estimating the variances.}
We have computed the MLE estimator for the means of the distributions subject to
the inequality constraints, up to the approximation of substituting the
individual sample variances instead of integrating them in the maximization
problem. Since the estimations of the means and variances
are inter-dependent, we may now need to reestimate the variances.

\begin{example}
Assume that we have samples $S_2 = \{0.1, 0.2\}$ for $Q_2$, and numerous samples
for $Q_1$ and $Q_3$ which convince us that $v_1 = 9$ and $v_3 = 8.5$ are very
good estimates for $q_1$ and $q_3$. We know that we must have $8.5
\leq v_2 \leq 9$ (we will probably choose $v_2 = 8.5$ given
$S_2$), but then our estimation of the variance of $Q_2$ should be
much higher than the sample variance $\widehat{\sigma}_2^2$ of $S_2$ in isolation.
\end{example}

We estimate the variance of each $X_i$ under the computed means $\mathbf{v}$
(and thus estimate the complete parameter $\boldsymbol{\theta}$) as the sample
variance relative to the computed mean $v_i$ of $X_i$ (instead of relative to
the sample mean). The solution thus obtained may not be optimal, as we have fixed and
optimized the means and variances separately rather than simultaneously.
Estimating how much this approach deviates from the true solution
is a challenge for future work.

\paragraph{Estimating the error and error decrease.}
The overall method now follows Section~\ref{sec:without} except that we follow the above\footnote{Note that this also changes the way of fitting distributions when computing the error decrease under possible additional samples.} to fit a family of distributions to the variables.

\section{Interpolation}
\label{sec:interpolation}
In some real-life scenarios, we may have a very large number of questions to ask the crowd; for instance, the number of possible compression ratios may be very high, almost continuous. In such cases, we may have many variables $X_i$ with no samples at all: those variables thus do
not appear in the optimization problem, so that we know nothing about them (except
that they satisfy the order constraints). However,
we could then perform \emph{interpolation} to estimate more precisely a large proportion of the
variables with a limited number of questions to the crowd.

In the general case where $E$ is an arbitrary set of inequalities, it
is hard to define how to interpolate a value for a variable with no samples.
We leave this general question to future work, and only focus on the
case where $E$ expresses the total order $\mu_1 \geq \cdots \geq
\mu_n$. For simplicity, up to renumbering indices, we assume that we have a
model for $X_1$ and $X_n$, namely $(\mu_1, \sigma_1^2)$  and $(\mu_n,
\sigma_n^2)$, and that we wish to derive a model for $X_k$, $1 \leq k \leq
n$, for which we have no samples.

\begin{example}
  \label{exa:interpol}
  If we estimate $v_1 = 8$ and $v_{5} = 4$, our best guess for $q_3$ in the absence
  of samples should be $v_3 = 6$. Likewise, our best guess for $q_4$ should be
  $v_4 = 5$.
\end{example}

\paragraph{Interpolating the mean.}

We interpolate the mean $\mu_k$ by a linear interpolation between $\mu_1$ and
$\mu_n$ according to the rank $k$, as presented in Example~\ref{exa:interpol}

\paragraph{Interpolating the variance.}

We want to interpolate $\sigma_k^2$ by combining both the variances
of $X_1$ and $X_n$, and the uncertainty arising from the interpolation itself:
the further away $k$ is from $1$ and $n$, the least certain
we are about~$\mu_k$.

To do so,
we consider that $\mu_k$ has been chosen by
picking $n-2$ random uniform values between $\mu_1$ and $\mu_n$ (the means $\mu_2,
\ldots, \mu_{n-1}$), sorting them, and choosing the $(k-1)$-th value to be $\mu_k$.
Now, this means that $\mu_k$ is the $(k-1)$-th \emph{order statistic} of $n-2$
uniform and independent random variables in $[\mu_1, \mu_n]$, so that it follows
a \emph{beta
distribution}~\cite{david2003order} whose variance has a closed form.

\begin{example}
  Pursuing Example~\ref{exa:interpol}, for $\mu_5 = 8$ and $\mu_9 = 4$, we
  estimate the variance on $\mu_7$ to be $4/5$ for this outcome (that of
  the adequate beta distribution).
\end{example}

We can thus estimate a variance for $X_k$ for fixed values $\mu_1$ and $\mu_n$:
those values are unknown, but can be sampled according to our model for
$X_1$ and $X_n$ to yield an overall variance for $X_k$.
We omit details for lack of space, and leave to future work the study of other
possible interpolation methods for variance.

\section{Conclusion and perspectives}
\label{sec:conclusion}
In this paper, we have studied the problem of learning numerical values from the
crowd, leveraging ordering constraints on those values to mitigate the
uncertainty on crowd answers. We have presented an abstract framework inspired
by~\cite{amsterdamer2013crowd} ignoring the order constraints, and presented an
approximate method to take those constraints into account, along with a way to
interpolate values for yet unsampled variables. We have identified further
challenges to be explored.

Our main direction for future work is to study more carefully the approximations
and design choices that we made, and to implement our approach to evaluate its
effectiveness. We plan to evaluate, over various datasets and objectives, the
importance of accounting for order constraints and performing interpolation, and
compare our approach to round-robin or random baselines, as well as ad-hoc strategies
for specific scenarios such as total orders.

\paragraph*{Acknowledgements.}
This work has been partially funded by the European Research Council under the FP7, ERC grant MoDaS, agreement 291071, and by the Israel Ministry of Science.

\bibliographystyle{abbrv}
\bibliography{main}

\end{document}